\documentclass[aps,prl,twocolumn,showpacs,floatfix,a4paper]{revtex4}
\usepackage{graphicx}
\usepackage[intlimits]{amsmath}
\usepackage{color}

\begin{document}

\title{Universal Voltage Fluctuations in Disordered Superconductors}

\author{A. Roy}
\author{Y. Wu}
\author{R. Berkovits}
\author{A. Frydman}

\affiliation{Department of Physics, Jack and Pearl Resnick Institute the and Institute of Nanotechnology and Advanced Materials,
  Bar-Ilan University, Ramat-Gan 52900, Israel}

\date{\today}

\begin{abstract} 

  The Aharonov-Casher effect is the analogue of the Aharonov-Bohm effect that applies to neutral particles carrying a magnetic moment.
  This can be manifested by vortices or fluxons flowing in trajectories that encompass an electric charge. These  have been predicted to
  result in a persistent voltage which fluctuates for different sample realizations. Here we show that disordered superconductors exhibit
  reproducible voltage fluctuation, antisymmetrical with respect to magnetic field,  as a function of various parameters such as
  magnetic field amplitude, field orientations and gate voltage. These results are interpreted as the vortex equivalent of the universal
  conductance fluctuations typical of mesoscopic disordered metallic systems.  We analyze the data in the framework of random matrix  theory and show that the fluctuation correlation functions and curvature distributions exhibit behavior which is consistent with Aharonov-Casher physics. The results demonstrate the quantum nature of the vortices in highly disordered superconductors both above and below $T_c$.

\end{abstract}


\maketitle

The Aharonov-Casher (AC) effect \cite{AC} is the dual effect to the Aharonov-Bohm (AB) effect \cite{AB}. While in the AB effect a charged particle quantum mechanical phase is affected by the electromagnetic vector potential, in the AC effect a neutral particle carrying a magnetic moment quantum mechanical phase is also affected by an "charge vector potential", even when no force is exerted on the particle.

\begin{figure}[htb]
\includegraphics[width= 8 cm,height=!]{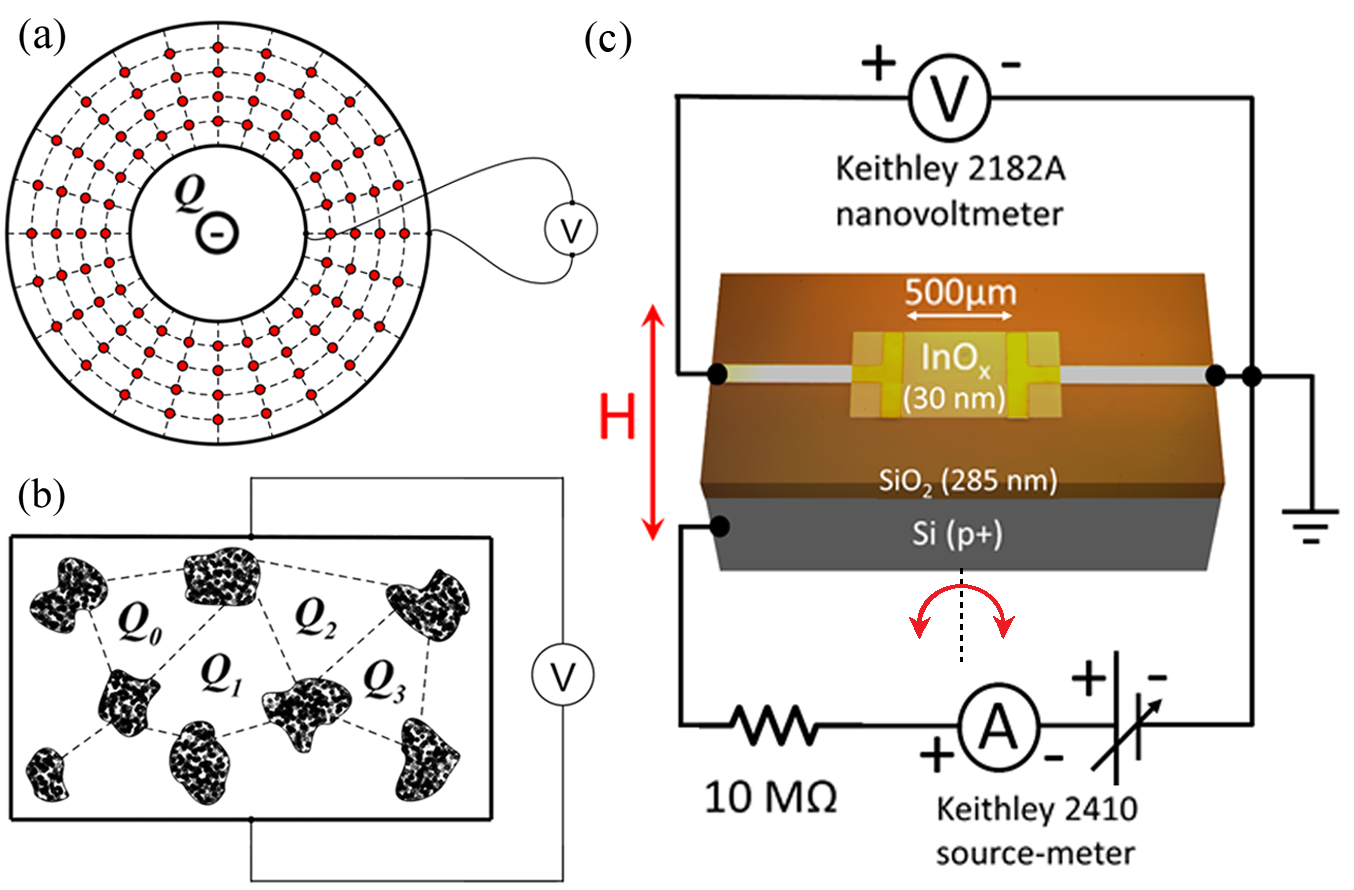}
\caption{\label{fig0}
  Sketches of systems expected to exhibit persistent voltage: (a) A 2D ring shaped
  Josephson array circling a charge $Q$ the persistent voltage is measured
  between the inner and outer edge. In order for a vortex to appear in the
  ring an external magnetic field perpendicular to the 2D plane must be applied.
  (b) A disordered 2D Josephson array composed of irregular placed and shaped
  superconducting islands. Depending on the magnetic field and
  on the charges trapped between the islands a reproducible random persistent
  voltage between two points on the edge of the sample is expected.(c) Schematic description of the sample geometry and measurement setup. The red arrow denotes the rotation axis in a magnetic field that is originally perpendicular to the substrate.}
\end{figure}

Magnetic vortices are an example of a neutral particle with a
magnetic moment and the AC effect was discussed for vortices in type II
superconductors \cite{reznik89,orlando91,zhu94}. A somewhat different realization of similar ideas is for a vortex (fluxon) in a two-dimensional (2D) Josephson-junctions arrays (see Fig. \ref{fig0}a). Although it carries no local magnetic flux, such a fluxon's phase is influenced by the charge encompassed by the array \cite{vanwees91} . Indeed oscillatory
behavior has been observed for transport measurements of such arrays \cite{elion93,pop12}.

The phenomenon of persistent currents is an illuminating demonstration
of the AB physics. Essentially when a ring encircles a magnetic flux $\Phi$ a persistent equilibrium current proportional to $\frac{\partial E}{\partial \Phi}$, where E is the energy, is predicted 
\cite{PCT1,PCT2,PCT3,PCT4,PCT5,PCT6} and measured \cite{PCE1,PCE2,PCE3,PCE4}.
van Wees \cite{vanwees91} realized that for the dual situation of a vortex in a 2D ring shaped array circling a charge $Q$ a persistent voltage between the inner
and external circumference of the ring proportional to 
$\frac{\partial E}{\partial Q}$ should appear.
This prediction has not been yet experimentally verified.

A different direction from the neat realizations described above is to
seek manifestations of AC physics in disordered samples (equivalent to mesoscopic disordered metals that exhibit universal conductance fluctuations (UCF)). Here, one gains in avoiding difficult preparation of the sample, but has much less control on the details (Fig. 1b). Specifically, a disordered 2D Josephson array composed of irregular placed and shaped superconducting islands is expected to exhibit AC physics manifested by reproducible voltage fluctuations for different sample realizations. 

In this letter we describe measurements of spontaneous voltage in amorphous indium oxide $(a-InO)$ films where the disorder is tuned so the samples are close to the superconductor-insulator transition (SIT). Indeed, it has been recently shown that the physics in the vicinity of this transition is determined by the Aharonov-Bohm-Casher effect \cite{Diamantini19}. The samples, despite being
morphologically uniform, have been shown to include \textquotedblleft emergent granularity" in the form of superconducting puddles embedded in an insulating matrix \cite{Kowal1994,Kowal2008,Shimshoni1998,Dubi2007,Imry2008,Trivedi1996,Ghosal2001,Bouadim2011,
poran2011,  shahar2012, sherman2014 Roy2018}, hence, they are perfect candidates for detection of AC effect signatures. For these films we find reproducible voltage fluctuations as a function of magnetic field amplitude, field orientation, and gate voltage. We analyze the results in terms of
random matrix theory and show that they exhibit universal features expected for the AC effect. 
\begin{figure}
\includegraphics[width=8.5cm,height=!]{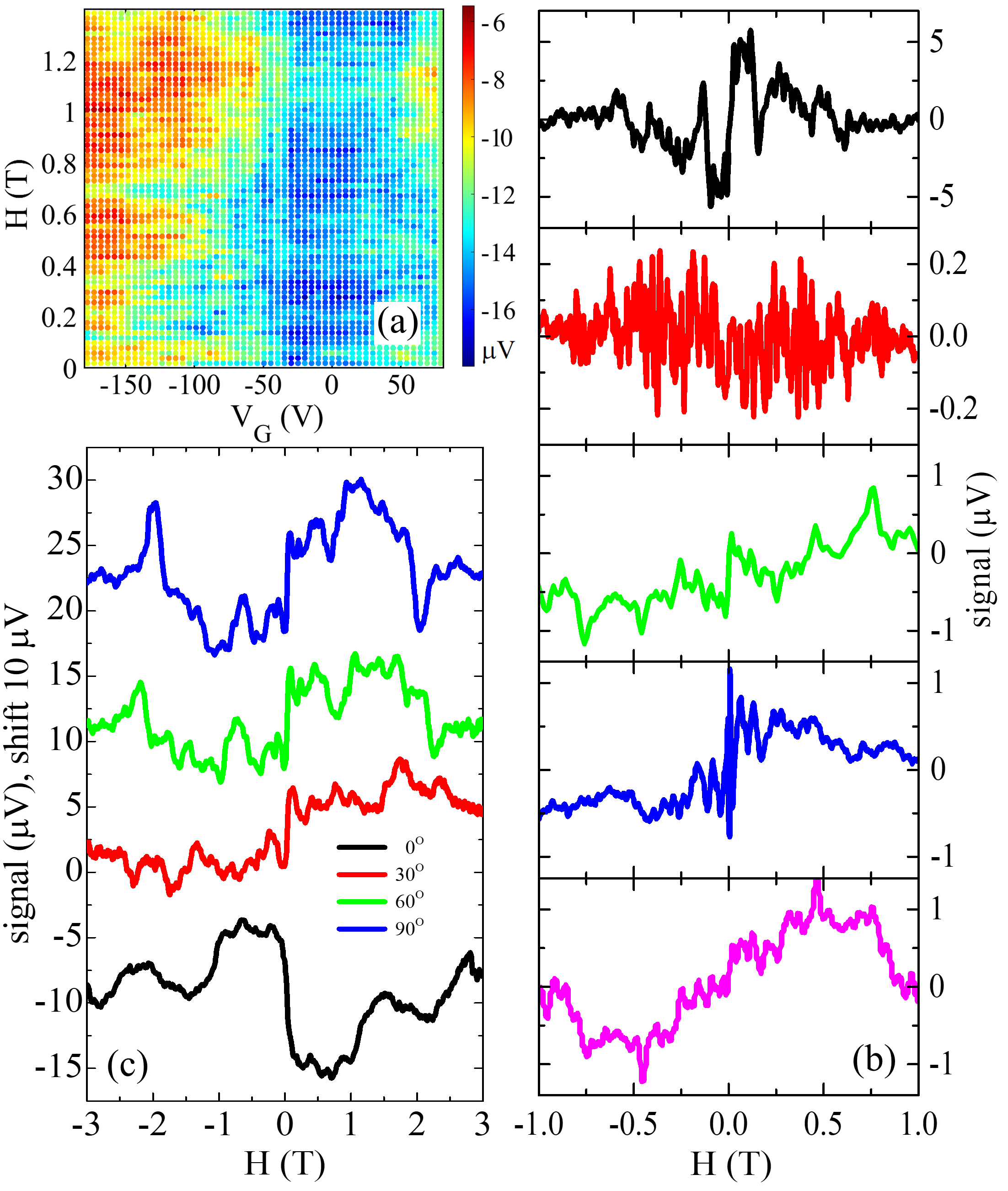}
\caption{\label{UVF}
  (a)  Spontaneous voltage as a function of gate voltage $V_G$ and out-of-plane magnetic field $H$ at 2K for a sample of $T_c=3.2K$. Quasi-periodicity arises both in response to magnetic field and enclosed charge on the islands, controlled by $V_G$.  (b) Spontaneous voltage as a function of out-of-plane magnetic field for a few $a-InO$ films. $V_g=0$. (c) Raw V(H) curves as a function of angle of the applied field relative to the film plane for one of our films. $V_g=0$. Voltage fluctuations w.r.t. the magnetic field are present even when the field is applied along the sample plane, indicating that processes other than the orbital effect are at work.}
\end{figure}

The studied samples were $a-InO$ films of thickness 30nm that were e-beam evaporated on MEMpax$\texttrademark$ borosilicate glass or gateable 
doped Si/SiO of thickness 0.4mm (sse Fig 1c). For these films the superconding coherence length is 10-30 nm \cite{shahar05} which places are films in the quasi-2D regime. The O$_2$ partial pressure during evaporation (in the range 1-8$\times$10$^{-5}$ Torr) determined the initial state of the sample, superconductor or insulator.  The results presented in this letter represent measurements
performed on 7 samples spanning the SIT with sheet resistance $R_{T=5K}$ ranging from 500$\Omega$ to 10 k$\Omega$. For more experimental details see supplemental material \cite{supp}.    

The natural parameter to vary  for obtaining different sample realizations for the AC effect is the gate voltage which controls the charges trapped between the islands and the various loops of vortex trajectories. However, for our 2D disordered samples on conventional substrates, obtaining detailed enough structure for analysis requires  unattainable large voltages. Fig. \ref{UVF}a shows that varying the gate voltage does result in voltage fluctuations. However, utilizing the variation of magnetic field as a driving parameter gives rise to much richer structure, hence we prefer this knob for generating AC voltage fluctuations. Depending on the magnetic field, which determines the vortex arrangement in the sample, a reproducible random persistent voltage between two points on the edge of the sample can be expected due to several different origins, the most obvious being the variation of the number of fluxons. Field may also change the superconducting properties of the islands as well as the charge distribution in the normal metal areas.

Fig. \ref{UVF}b depicts the voltage as a function of perpendicular magnetic field for a number of $a-InO$ film spanning the SIT. We note that for any realistic experimental setup one can not avoid some stray voltage,  $V_0$, which is present even at zero magnetic field. By subtracting this stray voltage, we obtain an anti-symmetric with magnetic field ($V(H)=-V(-H)$), reproducible structure.
The anti-symmetric nature of the fluctuations indicates that they originate from vortex motion. The structure is a true "fingerprint" of the sample microscopics in the sense that it is very reproducible for different magnetic field scans and scan rates (see supplemental \cite{supp}) of a single sample but it changes from sample to sample. Hence, we attribute these results to the AC equivalent of the UCF originating from the AB effect in disordered metallic  systems. 

It should be noted that unlike AB effect, AC physics does not depend directly on the magnetic flux penetrating the sample and thus
is not expected to be washed out when the orientation of the field is varied from perpendicular to parallel to the film. Indeed,
Fig. \ref{UVF}c, that shows the  V(H) curves (before anti-symmetrizing) as a function of magnetic field angle, demonstrates that though the voltage measurements depend on the magnetic field orientation, the  amplitude and the typical field scale of the fluctuations do not vary much as a function of field orientation.

\begin{figure}
\includegraphics[width=8cm,height=!]{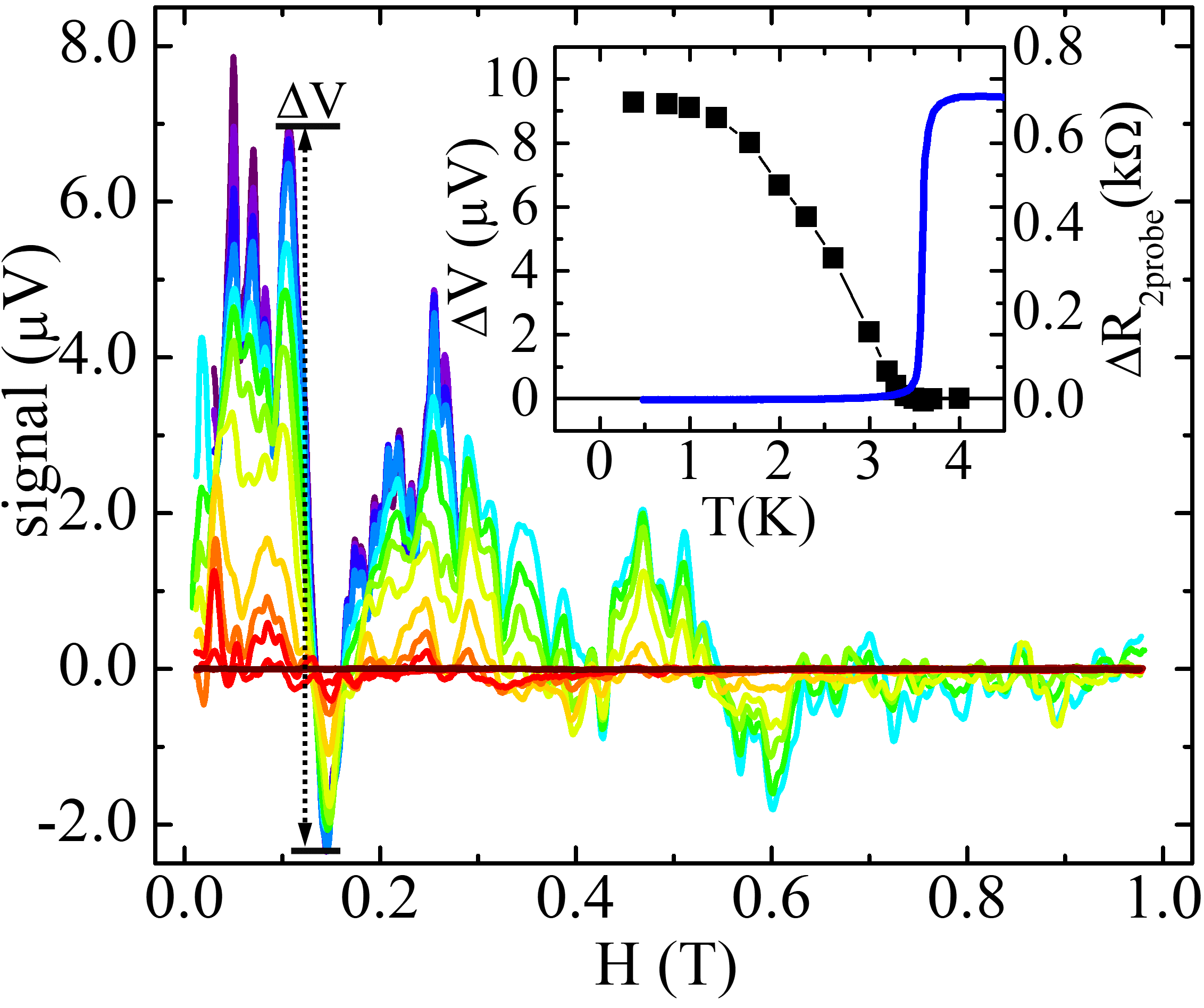}
\caption{\label{temp}
  Temperature dependent of the positive field regime of V-H curves for an $a-InO$ sample in the range T =0.39K(purple) to 4.00K(brown) (full curve at base temperature is shown in the supplemental \cite{supp}). $V_g=0$. Inset: Temperature dependence of the \textquoteleft amplitude' $\Delta V$ (black symbols) and resistance (blue line).}
\end{figure}

Fig \ref{temp} shows the monotonous suppression of the effect with growing temperature. As with the AB effect, decoherence, as result of e.g. temperature should reduce the observed effect. Nevertheless, the interpretation of effects of temperature on coherence should be done carefully, since it also effects the superconductivity of the grains composing the network and thus the vortices. A similar situation occurs also for the magnetic field.

Since in these conditions one can not model (nor control) the details of the sample one must look for global and statistical properties of the reproducable voltage in order to verify its origin. The experience of mesoscopic physics has taught us that even for disordered systems certain universal properties can be teased out of the data, which attest to the physics of the system \cite{imry02}. As in other cases of disordered mesoscopic systems \cite{gorkov65,efetov82,altshuler86,beenakker97} we assume that the universal part of the behavior of the system may be captured by a random matrix Hamiltonian. This assumption
works well for the description of diverse phenomena from correlations in the conductance at different magnetic fields \cite{efetov95} to the fluctuations in position and height of conductance peaks in the Coulomb blockade regime \cite{attias95,alhassid96,bruus96,sivan96}. 

The basic logic is similar here. We start with a rather general random matrix
Hamiltonian (For more details see supplemental material \cite{sup1})
describing a disordered 2D Josephson array composed of irregular placed and shaped superconducting islands. Depending on the magnetic field and on the charges trapped between the islands a reproducible random persistent voltage between two points on the edge of the sample is expected. Keeping track of all the influences of magnetic field on the vortex motion is a rather herculean task. Nevertheless, we can exploit the theory developed for correlations \cite{szafer93,simons93,simons93a,beenakker94}
and curvature distribution \cite{zak93,vonoppen94,fyodorov95,fyodorov95a,canali96,yurkevich97,avishai97,basu98} of the spectral response to external parameters which show universal behavior of the derivative of the energies of the system with respect to an external parameter. Specifically, for the
$i$-th eigenvalue $\epsilon_i(x)$ (where $x$ is the value of the external parameter) of the Hamiltonian one may define a ``velocity'' $j_i= \partial_x \epsilon_i(x)/\delta$ (where  $\delta$ is the mean level spacing, i.e. a system dependent parameter) and a ``curvature'' $K_i= \partial_x \j_i(x)$ characterizing the response of the spectrum to the external parameter. This corresponds to the identification of the persistent voltage with the
current and its derivative with the curvature.
The correlation
\begin{equation}\label{cor}
C_i(\delta x) = \langle j_i(x) j_i(x+\delta x) \rangle,
\end{equation}
where $\langle \ldots \rangle$ is an average over different systems and
ranges of $x$. After a renormalization of 
the parameter $X=\sqrt{C_i(0)} x$, a universal behavior emerges \cite{szafer93,simons93,simons93a,beenakker94}
where
\begin{equation}\label{coru}
C_i(\delta X) = -\frac{2}{(\beta \pi^2 \delta X)^2},
\end{equation}
as long as $\delta X$ is larger than some non-universal value
and $\beta=1$ for the orthogonal ensemble (GOE) and $\beta=2$ for the unitary
one (GUE). It is important to note that since $x$ depends on a non-universal
parameter of the system $\delta$ , which is hard to obtain, one can determine
$X$ only up to a factor.
This leads to an unusual correlation curve since the correlation should be maximum
at $\delta X=0$, and approaches zero from below an large $\delta X$, which
means that there is a negative minimum of the correlation at some intermediate
value of $\delta X$. For the curvature, a universal distribution is expected
\cite{zak93,vonoppen94,fyodorov95,fyodorov95a}.
Defining $k=K_i/|\langle K_i \rangle|$, one obtains the distribution

\begin{equation}\label{dis}
  P(k)=\frac{A_{\beta}}{(1+k^2)^{(2+\beta)/2}},
\end{equation}
where $A_1=1$,$A_2=(4/\pi)$.
Corrections to this distribution, which are especially notable
at low values of $k$ as result of non-universal features have also been
discussed \cite{yurkevich97,basu98}. Since here one normalizes the curvature
by its absolute averaged value, the dependence on $\delta$ disappears.
The negative minimum of the correlation as well as the distribution
which has no adjustable fit parameters are a ''smoking
gun'' evidence for the random matrix description of the physics of the
system as well as the identification of the fluctuation in voltage with the
derivative of the energy, in agreement with the Ahronov-Casher picture.

\begin{figure}
\includegraphics[width=7.5cm,height=!]{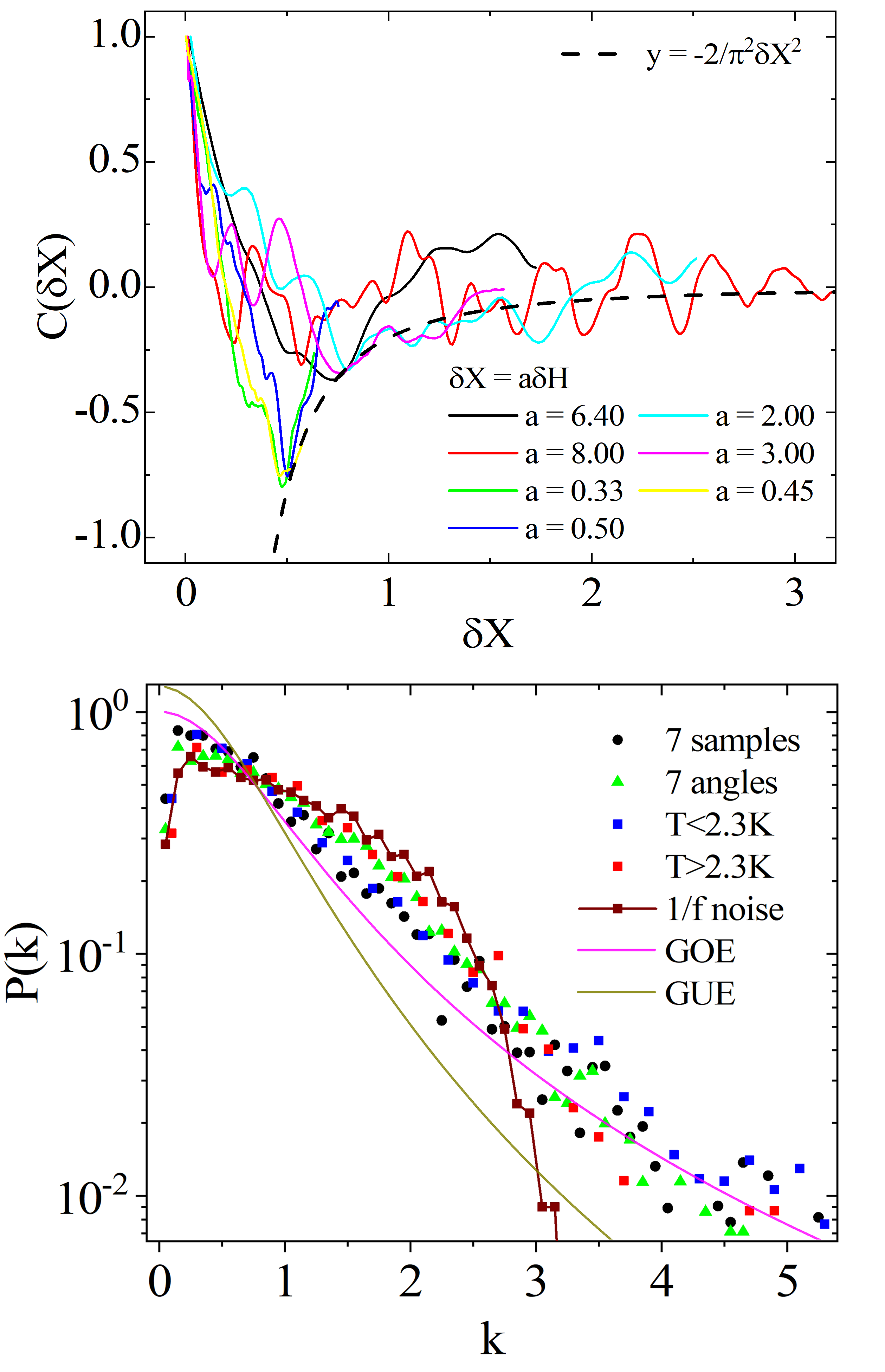}
\caption{\label{theory}
  Top: The correlation $C(\delta X)$ for seven different samples where $\delta X =a  \delta H$ and $a$ is a sample dependent rescaling
  constant. The dashed curve corresponds to $-c/(\delta X)^2$ , with $c=2/\pi^2$ (GOE).
As discussed in the text plotting the GUE curve ($c= 1/\pi^2$)  will result with same figure with rescaled values of $a$ by half.
  Bottom: The distribution $P(k)$ aggregated over seven different samples, as well as the distribution for seven different angles of the external
  magnetic field, and for low $T<2.3K$ and high $T>2.3K$ temperatures ($T_c$, defined at the temperature at which the resistance drops to 10\% of its normal value, equals 2.3K) . For comparison, the distribution for a numerical sequence
  of $1/f$ noise is presented. The GOE ($\beta=1$) and GUE ($\beta=2$) distributions are plotted. }
\end{figure}

Here we analyze the results of seven different samples, for one of them, we analyze the data for $7$ different angles of the magnetic field with respect to the sample plain, and for another sample at a range of temperatures between $T=0.42K$ and $3.26K$.
After subtracting the stray voltage background $V_0$ we antisymmetrize the data in the following way:
\begin{equation}\label{as}
  V_{AS}(H)=\frac{V(H)-V(-H)}{2}-\bar V,
\end{equation}
where $\bar V$ is the average over $(V(H)-V(-H))/2$ for the whole range of
measurement so $\langle{V_{AS}(H)}=0\rangle$.

In order to substantiate these observation we calculate the correlation:
\begin{equation}\label{cor}
  C(\delta H)=\frac{\langle V_{AS}(H+\delta H)V_{AS}(H)\rangle_H}
  {\langle V^2_{AS}(H)\rangle_H},
\end{equation}
where $\langle \ldots \rangle_H$ denotes an average over the available
values of the magnetic fields. The samples show a peak in the correlation which then turns into negative values and then returns to values around zero. Rescalling the correlations such that $\delta X = a \delta H$, with $a$ a sample specific constant  results in
a very similar correlation curve for all samples (see Fig. \ref{theory}a). It is important to note that since $a$ depends on a microscopic scale which is unknown to us, here it is
not possible to distinguish between GOE and GUE ($\beta=1$ ,$\beta=2$). Nevertheless,
the width of the correlation for all samples is similar after rescaling, and the correlation follows the expected $\propto (\delta X)^{-2}$ behavior.

In Fig. \ref{theory}b the experimental distribution for several different ensembles are plotted. Since here we can avoid sample specific fit parameters by dividing the curvature by its sample average over different magnetic fields, we can pool together data from several samples, angles of the external magnetic field and temperatures. All of these distributions have several common features. They follow quite closely the GOE distribution, especially if one compares distribution from numerical generated $1/f$ noise sequence. The deviations are most pronounced for small values of $k$ as has been been expected from non-universal corrections to the distribution \cite{yurkevich97,basu98}. It is also notable that the measurement of $7$ different angles treated as independent samples yields similar results as for $7$ different samples. This lends further support to the fluxon interpretation which is not based on orbital effects. 

Although as seen in Fig. \ref{temp} temperature suppresses the amplitude of the voltage fluctuation it hardly wipes out the curvature distribution although at temperatures above $T>2.3K$ the global superconductance is strongly suppressed. This hints that there is no need for a global coherent superconductivity, and remnant local superconductivity suffices. Similarily, we obtain repeatable  voltage structure for samples in the insulating phase though these have not been analyzed here. Indeed, local superconductivity, manifested by a finite gap, was observed both above $T_c$ \cite{sacepe} and in the insulating phase \cite{sherman} of $a-InO$ films.   This may also 
address another puzzle. A magnetic field breaks time reversal symmetry and therefore one expects the appropriate random matrix ensemble to be GUE. Here, the distribution is GOE, which can be understood if the mechanism for the voltage fluctuation is short ranged and
therefore time reversal symmetry is not broken on that length scale.
 
In summary, granular samples around the superconducting transition  are promising candidates for experimental studies of  additional features of random matrix theory. In our disordered superconductors, that incorporate "electronic granularity", the correlations and curvature of the voltage fluctuations fit much better the expectation for a system that follows the universal features of random matrix theory than those of non-correlated noise (e.g. $1/f$ noise). Nevertheless, it would be strange if non-universal features would not appear in realistic sample and open venues for further study. Although some features such as level spacings and wave function properties have been extensively studied experimentally in systems such as quantum dots \cite{beenakker97}, features such as level velocity and curvature to the best of our knowledge have not. The identification of the voltages fluctuations as as an indication to the AC effect emphasize the quantum nature of the vortices close to the SIT, below and above $T_c$.

\begin{acknowledgments}
We are grateful to I. Volotsenko for technical help and to K. Behnia, E. Shimsoni, N. Trivedi and V. Vinokur for useful discussions. This research was supported by the Israel science foundation, grant No. 783/17.
\end{acknowledgments}

\section{Supplementary Material}
\subsection{1.	Details of samples and measurements}
For studying universal voltage fluctuations (UVF) in the disordered superconductors, we fabricated amorphous InO$_x$x films on MEMpax\texttrademark   borosilicate glass or doped Si/SiO$_2$ substrates by photolithography and e-beam vapour deposition. As shown in Fig.1c, the geometry of the samples was a square of 500$\times$500 $\mu$m and 30nm thickness. The sheet resistance ranged between 500$\Omega$ - 10k$\Omega$ as a result of partial Oxygen pressure of $1\times 10^{-5} - 8\times 10^{-5}$ Torr during deposition. For the Si/SiO$_2$ substrates we grounded one side of the InO$_x$ and applied a variable dc gate voltage with the doped Si utilized as a gate electrode (see Fig.1c). 
Most measurements were performed in a helium-3 cryostat with base temperature of 300mK. One gated sample was measured in a pumped helium-4 system with base temperature of 1.4K. The system had a rotatable sample mount for studying the dependence of the measured signal on the angle of the applied field. For reducing external noise, our voltage probes were connected directly to the nanovoltmeter without intermediate joints thus minimizing spurious thermoelectric voltages.
The V(H) curves were obtained by sweeping a magnetic field in the range of $\pm$4T or a gate voltage in the range $\pm$180V (depending on the sample and substrate) for various temperatures. The voltage was measured using a Keithley 2182A nanovoltmeter with sensitivity of 1nV. The gate voltage was applied by a Keithley 2410 source-meter through a 10M$\Omega$ resistor. Current in the gate circuit was monitored to exclude the possibility of gate leakage.

\subsection{2.	Noise level}
\begin{figure}
\includegraphics[width=7.5cm,height=!]{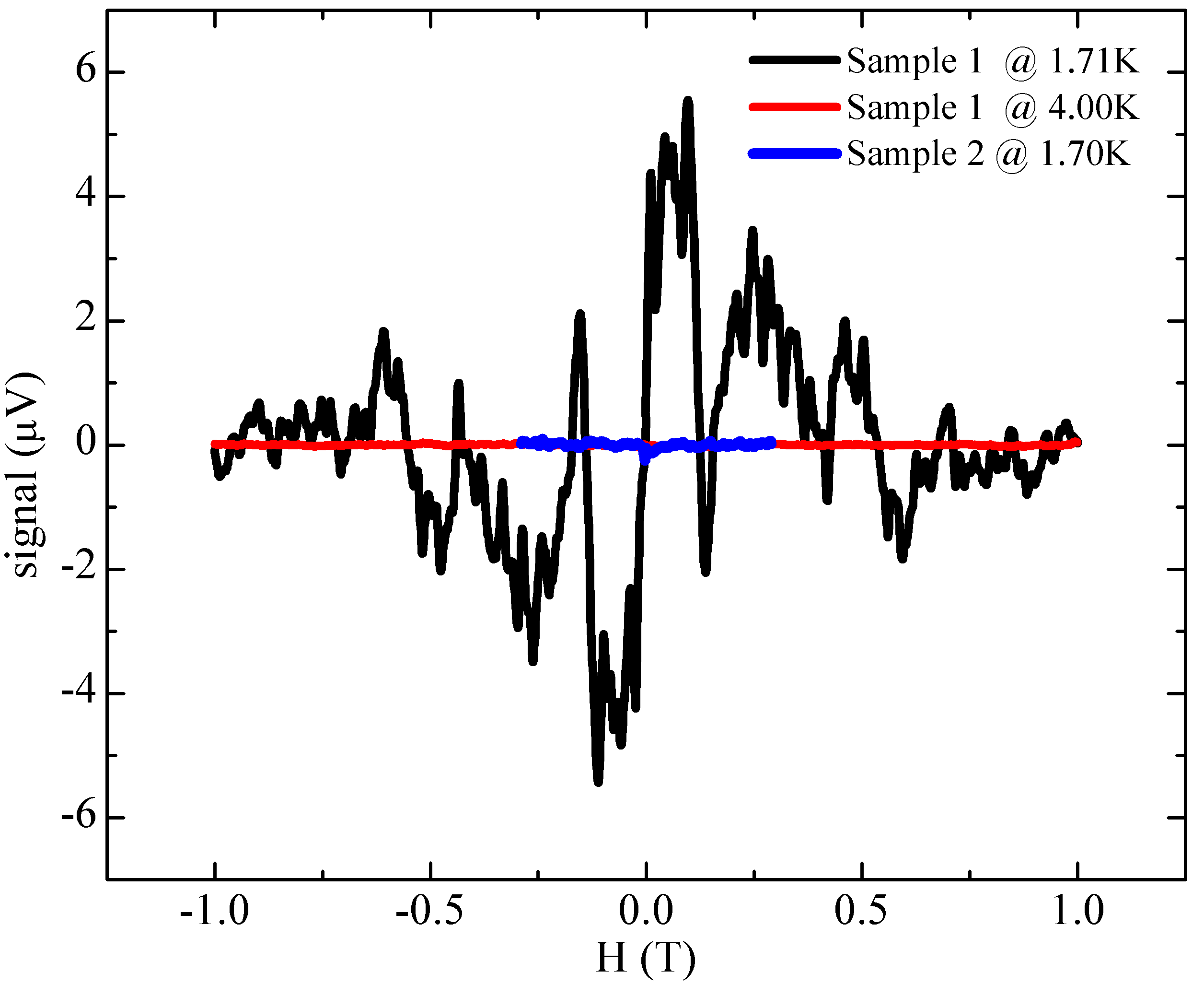}
\caption{\label{S1}Magnetic field responses of two samples with sheet resistances of 800$\Omega$ (blue curve) and 1500$\Omega$ (black and red curves). The black curve is for the sample of Fig.\ref{temp} demonstrating the anti-symmetric nature of the curve. A constant dc background voltage has been subtracted from each signal.}
\end{figure}

For determining the noise floor of our measurement setup we compare the responses of a sample at two temperatures, below and above the superconducting T$_c$ (3K). At 1.7K where the sample is superconducting, the antisymmetric voltage fluctuations have an amplitude of $\sim$6 $\mu$V (Fig.\ref{S1} black curve). In the absence of superconductivity (at 4K), the signal drops to the noise floor of our setup which is slightly higher than 10nV (Fig.\ref{S1} red curve). We also compare this sample (R$\mu$ = 1500$\Omega$ at 5K, labelled as Sample1) with another sample deep inside the superconducting regime (R$\mu$ = 800$\Omega$ at 5K, labelled as Sample2). In the case of the latter, the signal is not antisymmetric, and fluctuates in the range of $\pm$50nV (Fig.\ref{S1} blue curve). This is due to the fact that a sample deep in the superconductor does not exhibit the rich microscopic disorder caused by superconducting puddles in an insulating matrix which characterizes samples close to the SIT. 

\subsection{3.	Effect of field sweep}
\begin{figure}
\includegraphics[width=7.5cm,height=!]{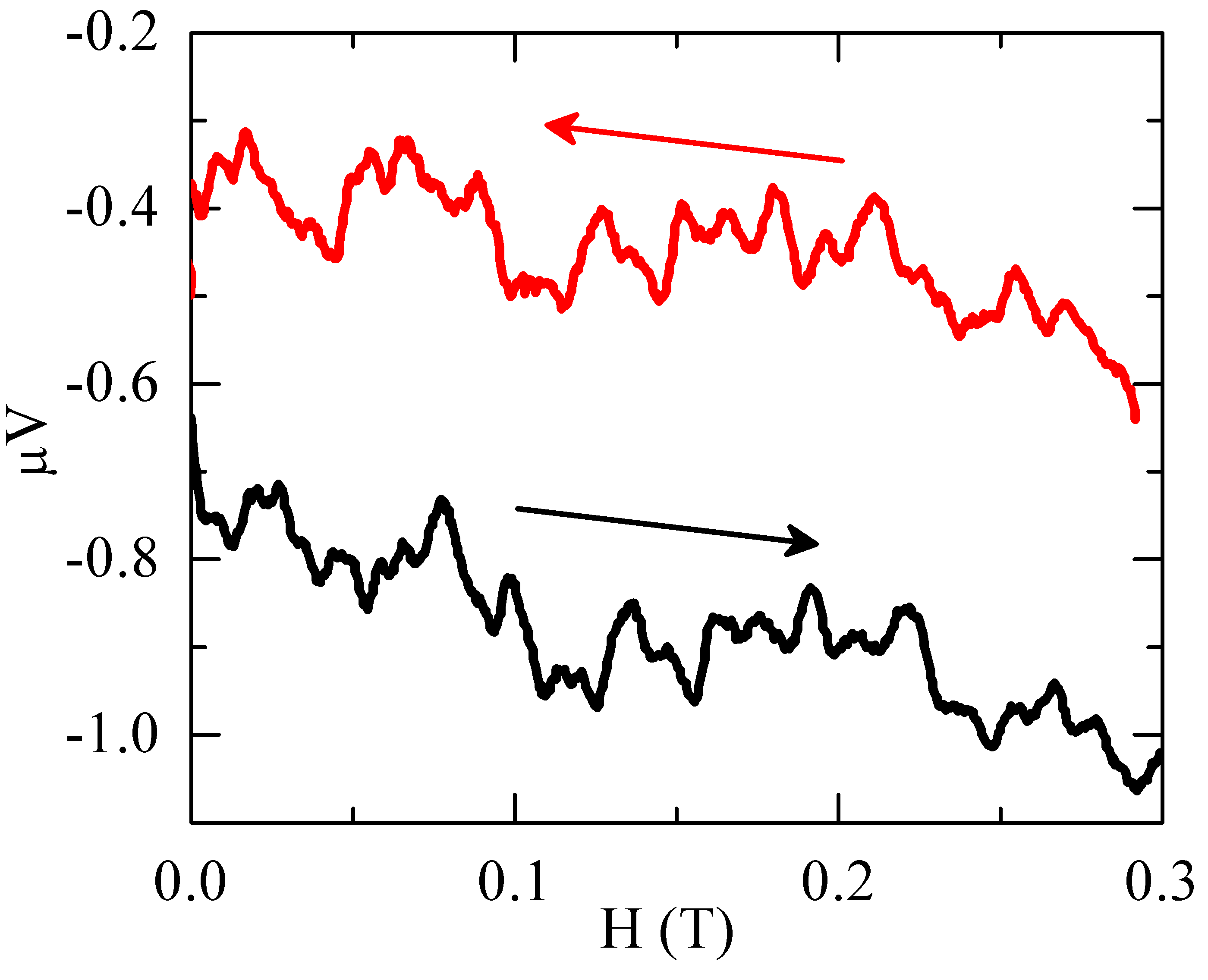}
\caption{\label{S2} As-measured voltage signal from a typical sample for opposite field sweep directions, indicated by arrows. A constant difference of $\sim$200nV is present due to the induced voltages being of opposite sign.}
\end{figure}

A magnetic field sweep adds an additional dc voltage due to inductive pickups from the measurement leads. At a constant field sweeping rate, this voltage appears as a constant background voltage over which the voltage fluctuations are superposed. The fluctuations themselves are independent of the sweep direction and antisymmetric w.r.t. the field. This is illustrated in the Fig.\ref{S2} showing the as-measured signal for opposite sweep directions. For the fluctuation analysis, the background is removed by numerical antisymmetrization.

The rate of sweep has minimal effect on the observed fluctuations, except for the fact that small features may sometimes get washed out for a sweep that is too fast, due to the built-in filters of the measuring instrument (Fig.S3). For a different sample, the voltage signal during a sweep was compared with a measurement protocol where the field sweep was halted for each measurement and sufficient time (~60sec) was allowed for the sample to relax (Fig.S4). This eliminates the possibility of artefacts that arise purely due to the magnetic field changing with time.

\begin{figure}
\includegraphics[width=7.5cm,height=!]{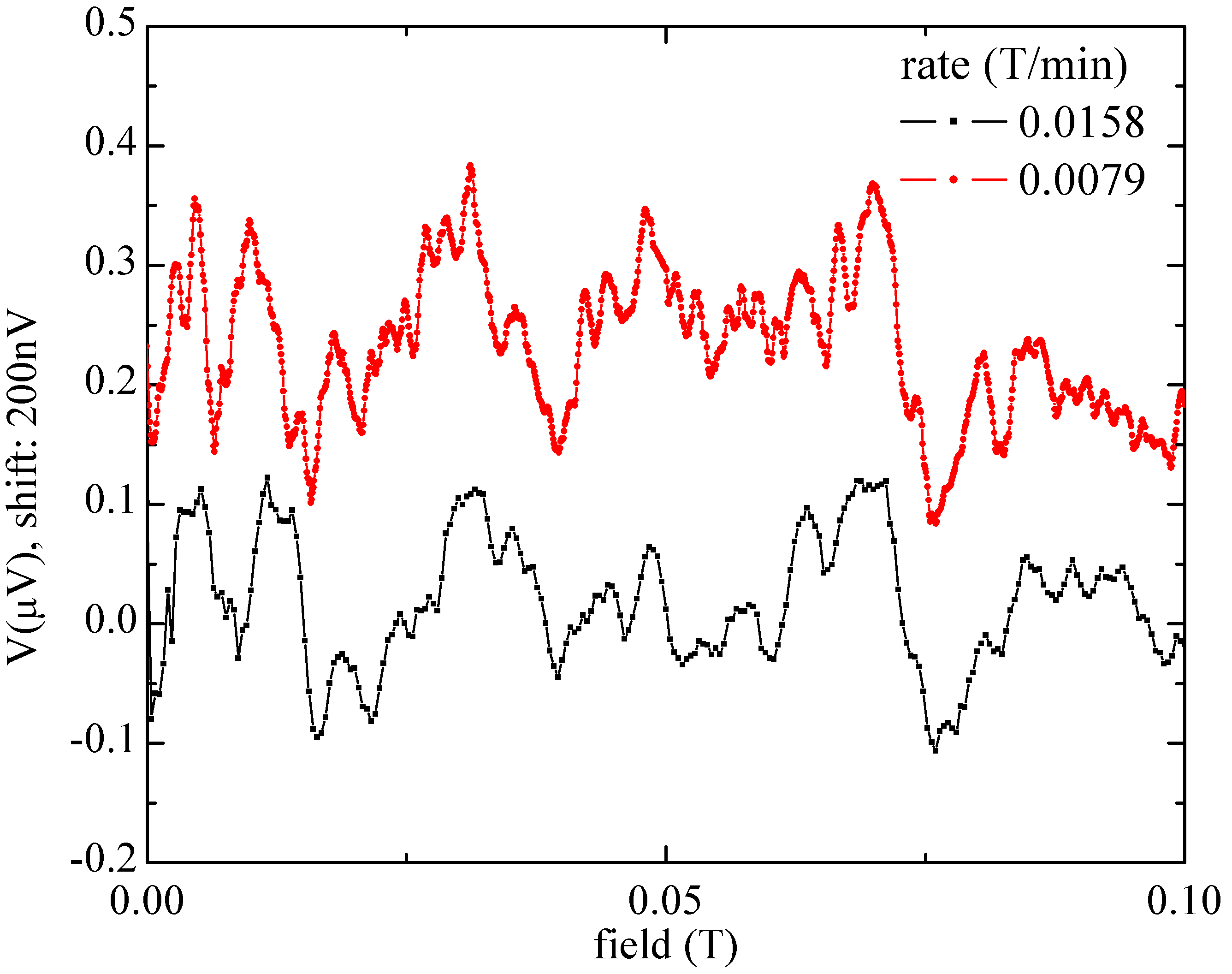}
\caption{\label{S3} Raw Signal for two different sweep rates in the same direction. A constant dc background voltage has been subtracted from the as-measured signals, and the red curve has been shifted up by 200nV for clarity. The faster sweep (black curve) shows fewer features, possibly due to the action of the built-in filters inside the measuring instrument.}
\end{figure}

\begin{figure}
\includegraphics[width=7.5cm,height=!]{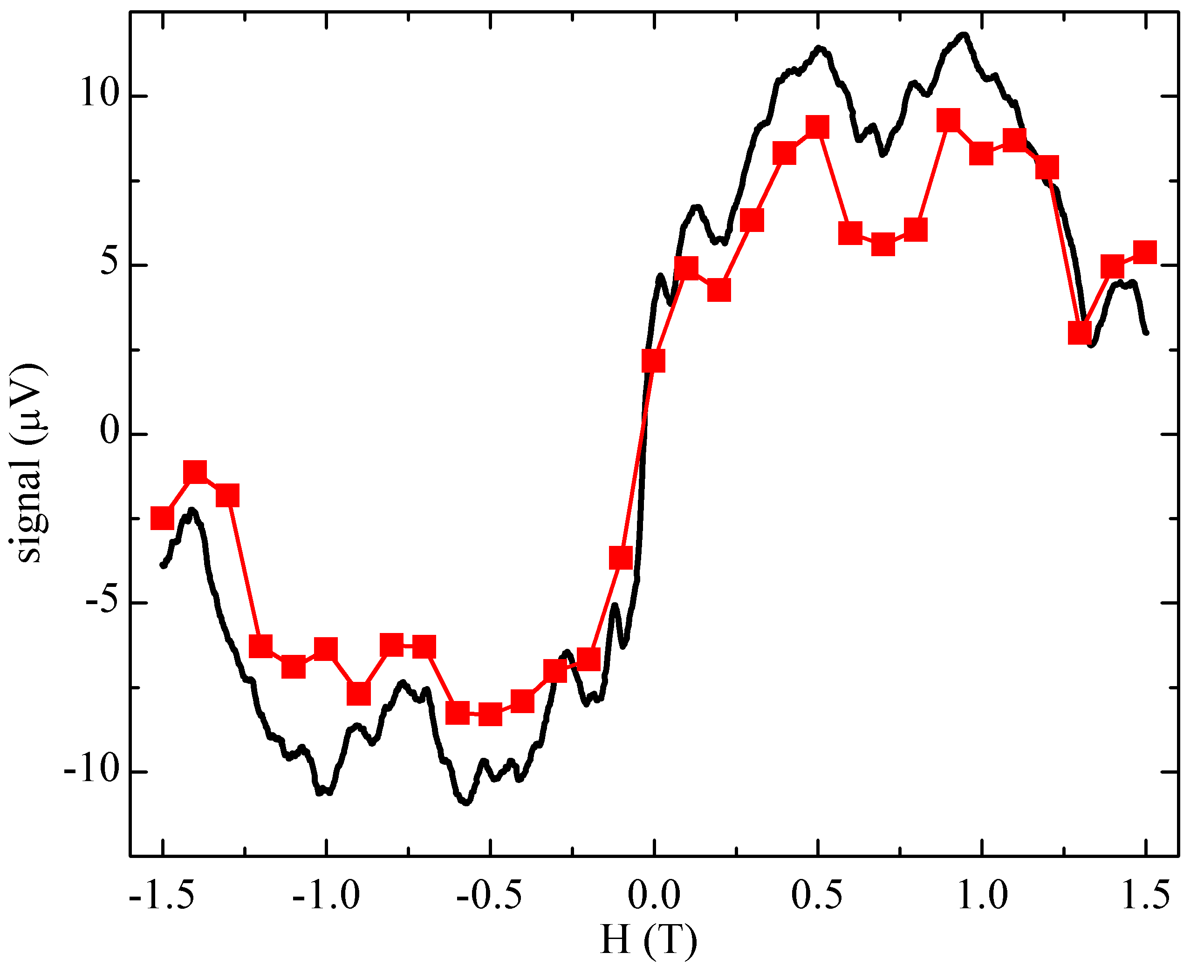}
\caption{\label{S4} Comparison of measurements by two protocols. Black line: As-measured signal during a field sweep (mean subtracted). Red symbols: As-measured signal under the same conditions in using a relaxation + acquisition protocol (mean subtracted), data being collected after waiting $\sim$ 60 sec at each field value. }
\end{figure}

\section (4.   Random Matrix Theory and Ahronov Casher Effect)

As suggested by Wigner in the context of modeling the spectra of heavy nuclei,
when one cannot fully describe the Hamiltonian of a complex system, some
properties which depend on the symmetry of the system
nevertheless may be captured. By representing the Hamiltonian as a
random matrix and studying the emergent statistics of the eigenvalues and
eigenvectors of an appropriate random matrix ensemble
\cite{wigner55,mehta04}. The distribution of the level spacings between
adjacent eigenvalues of a
is the canonical example for identifying universal random matrix
signature in experimental data, but many other features have been studied.

For associating the statistical properties of the voltage fluctuations and
the properties of random matrices, one should connect statistical
properties of the
eigenvalues of a random matrix to the voltage. For a simply connected
geometry the experiment the voltage
between two points on the edge of the sample is associated
to the derivative of the systems energy with respect to the charge
trapped in the void. For a complex multiple connected system of the type
measured in the experiment, the dependency is much more complicated, akin
to the difference between the AB response of a ring to a network. 
Nevertheless, the basic connection between the voltage and the derivative
of the energy with respect to an external parameter holds.

\begin{figure}
\includegraphics[width=8cm,height=!]{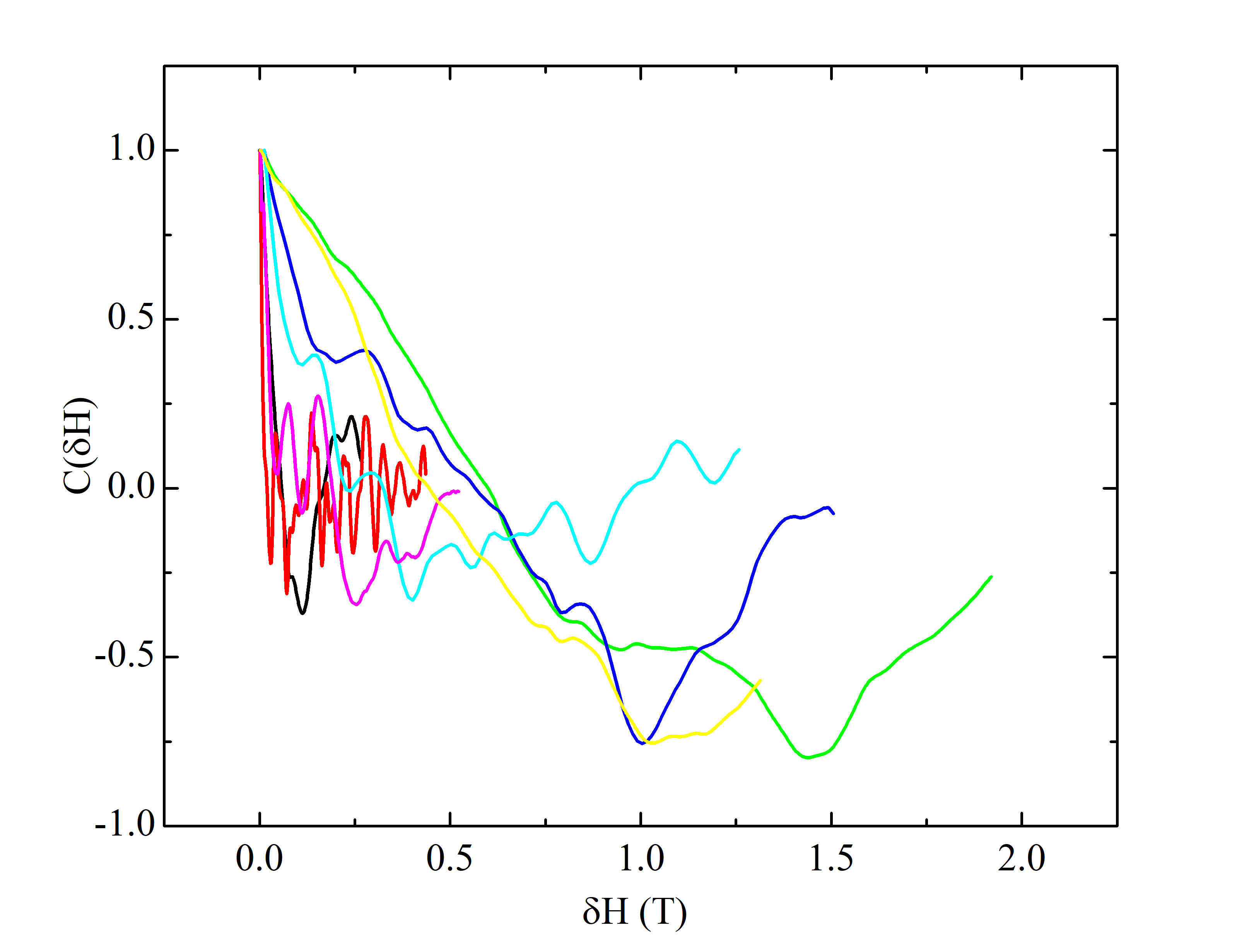}
\caption{\label{correlation}
  The correlation $C(\delta H)$, where $H$ is given in Tesla, for seven different samples, prior to the renormalization leading to Fig. 4 in the main text.  }
\end{figure}

Thus, we should look for statistical properties of derivatives of the
eigenvalues of random matrix ensemble with respect to an external
parameter. As detailed in the main text, we
concentrate on two such properties. 
Correlations between the parametric derivatives of the eigenvalues at different
values of the parameter \cite{szafer93,simons93,simons93a,beenakker94}
and the distribution of the second derivative with respect to the external
parameter
\cite{zak93,vonoppen94,fyodorov95,fyodorov95a,canali96,yurkevich97,avishai97,basu98}
normalized to its absolute average. The derivative and second derivative
of the eigenvalue with respect to an external parameter, $x$, for an
eigenvalue $\epsilon_i(x)$ of the matrix
termed the  velocity and the curvature,
are $j= \partial_x \epsilon(x)/\delta$ and
$K= \partial_x \j(x)= \partial^2_x \epsilon(x)/\delta$
respectively. Note that the mean level spacing $\delta$ is non-universal.
The measured voltage $V_{AS}(H)$ or $V_{AS}(V_g)$
are proportional to the velocity, $j$,
while $\partial V_{AS}(H)/\partial H$ or
$\partial V_{AS}(V_g)/\partial V_g$ correspond
to the curvature $K$. The proportionality depends on system specific properties,
which are undetermined by the measurement. The correlation
$C(\delta H)$, as defined in Eq. (5) in the main text, is
calculated for each sample and averaged over a range of magnetic fields
$H=0  \ldots H_{max}-\delta H_{max}$ and is depicted in Fig. \ref{correlation}.
Now one remains with associating $\delta H$ to the rescaled external parameter
$\delta X$. With no information on the microscopic's of the system it is
not possible to estimate the effective level spacing $\delta$ in the
corresponding random matrix model, and therefore no way to evaluate
proportionality $\delta X \propto  \delta H$. Nevertheless, from
examining Fig. \ref{correlation} it is clear that the main feature of
the universal
correlation, i.e., a distinct minimum at intermediate values of $\delta H$,
then approaching zero from below at higher values, is apparent for all samples.
Therefore, a constant $a$, corresponding to $\delta X = a \delta H$ is chosen
for each sample so the curves fall on each other. Of course, the constant is
arbitrary and therefore can not be used for example to determine whether the
correlation fits GOE or GUE as discussed in the main text.

Fortunately, For the curvature, we have the distribution as function of the
normalized curvature $k=K_i/|\langle K \rangle|$
\cite{zak93,vonoppen94,fyodorov95,fyodorov95a} , where $\langle K \rangle$ is
averaged over the different values of the external magnetic field $H$.
By normalizing the curvature the non-universal quantity, $\delta$, drops
out and therefore, as detailed in the main text, it is possible to
determine the symmetry (GOE) of the system.

\end{document}